\documentclass[baaa]{baaa}
\usepackage[pdftex]{hyperref}
\usepackage{subfigure}
\usepackage{natbib}
\usepackage{helvet,soul}
\usepackage[font=small]{caption}

\contriblanguage{1}

\contribtype{1}

\thematicarea{6}

\title{Studying the molecular gas towards a bright rimmed cloud at the infrared dust bubble N30}

\titlerunning{A bright rimmed cloud at bubble N30}

\author{A. Solernó\inst{1}, M. B. Areal\inst{1}, \& S. Paron\inst{1}}
\authorrunning{Solernó et al.}

\contact{sparon@iafe.uba.ar}

\institute{
Insituto de Astronomía y Física del Espacio, CONICET--UBA, Argentina 
}

\resumen{
Se presenta un estudio del gas molecular hacia una nube de borde brillante ubicada hacia el norte de la 
burbuja infrarroja de polvo N30.
Utilizando la emisión de la línea J=3--2 del $^{12}$CO, $^{13}$CO, y C$^{18}$O, junto a datos del infrarrojo y del 
continuo de radio se realizó una caracterización
de la burbuja y de la nube molecular relacionada. Adicionalmente se presenta un análisis del comportamiento de la
relación de abundancia $^{13}$CO/C$^{18}$O hacia la nube de borde brillante y se buscan indicios de formación estelar
reciente en la región.
}

\abstract{
We present a study on the molecular gas towards a bright-rimmed cloud located to the north of the infrared dust bubble N30. Using the emission from the $^{12}$CO, $^{13}$CO, and C$^{18}$O J=3--2 line, together with infrared and radio continuum data, we characterized the bubble and the related molecular cloud. In addition, we show an analysis of the behaviour of the abundance ratio $^{13}$CO/C$^{18}$O towards the bright-rimmed cloud, and we search for clues on recent star-formation.
}


\begin{document}

\maketitle

\section{Introduction}

The H\textsc{ii} regions identified as infrared dust bubbles in \citet{church06}, are promising objects to study the interplay between the ionization/photodissociation and the cold molecular gas. Their study can provide us an insight into the properties of the interstellar medium (ISM) in which they are expanding and the possible triggered star formation mechanisms that may occur in their surroundings. In particular, we have studied bubble N29 and its secondary inner bubble N30, centred on the galactic coordinates l=23.10, b=+0.58, at a distance of 2.2 kpc, with a systemic velocity {\it v}$_{LSR}$ of about 30 km s$^{-1}$ \citep{bea10,deha10}.

\begin{figure}[!t]
    \centering
    \includegraphics[width=0.47\textwidth]{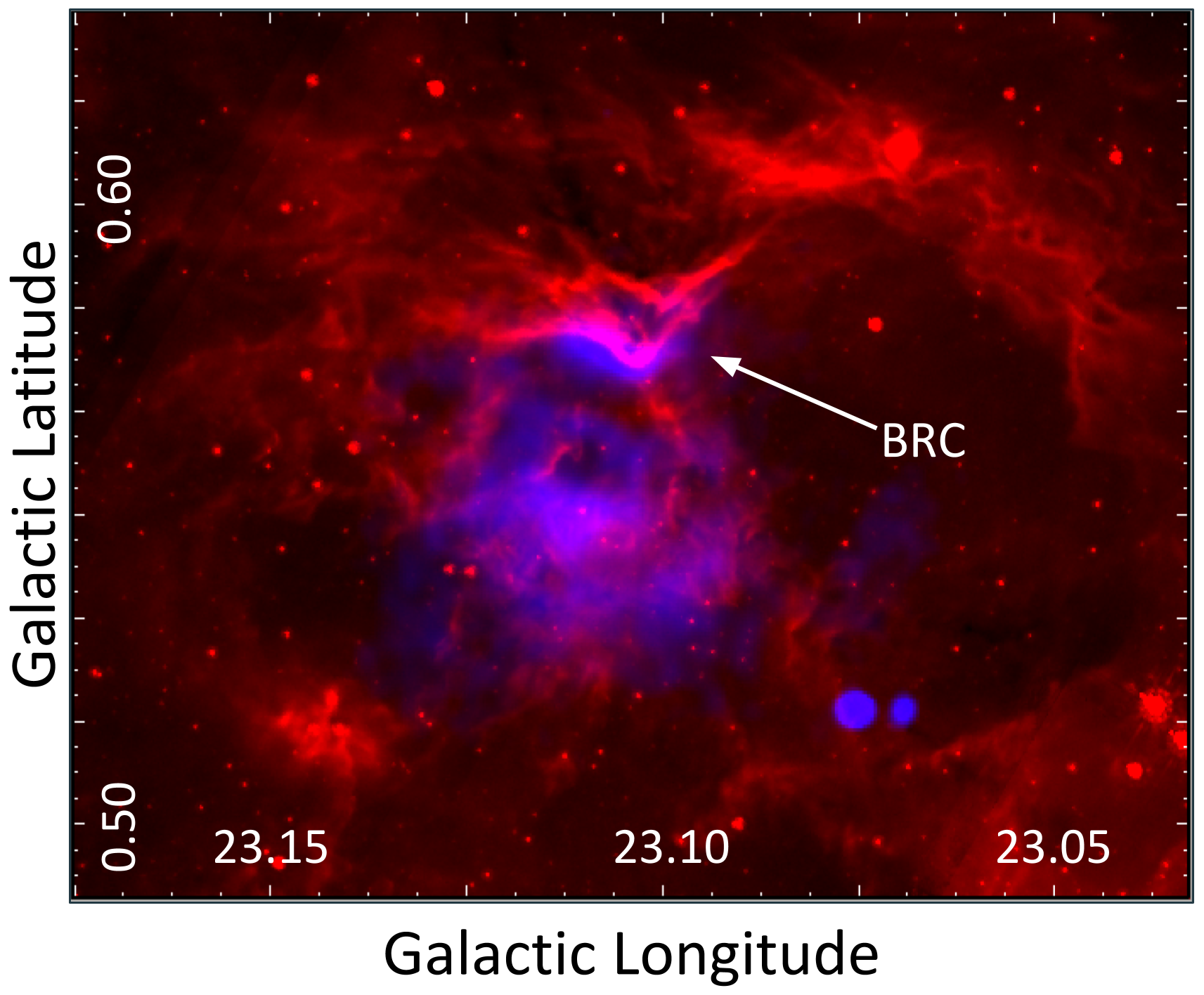}
    \caption{GLIMPSE/{\it Spitzer} emission at 8~$\mu$m (red) and the radio continuum emission at 20~cm from the MAGPIS (blue) towards bubble N30. The presence of a BRC is indicated.}
    \label{fuente}
\end{figure}

Figure\,\ref{fuente} shows an image of the studied region displaying the 8~$\mu$m emission extracted from the GLIMPSE/{\it Spitzer} survey\footnote{https://irsa.ipac.caltech.edu/data/SPITZER/GLIMPSE/} (red) and the radio continuum emission at 20 cm 
obtained from the MAGPIS\footnote{https://third.ucllnl.org/gps/} (blue). In particular the emission at 8~$\mu$m allows us to identify a pronounced curvature towards the upper edge of bubble N30, where the ionized gas, traced by the emission at 20 cm, is very likely stalling, suggesting that the ionization and photodissociation fronts of the H{\sc ii} 
region are interacting with a dense molecular clump. This type of structure is known as bright rimmed clouds (BRCs) and it is usually associated with the star formation process known as radiative driven implossion (RDI) (e.g. \citealt{morgan09}). This BRC, not previously studied, is an interesting feature to be analyzed through the emission of molecular lines.

\section{Molecular data}

We used molecular data of the $^{12}$CO, $^{13}$CO and C$^{18}$O J=3--2 emission obtained from the public database of the 15~m James Clerk Maxwell Telescope. The $^{12}$CO data correspond to the observation program M08AH03A, while the data of the $^{13}$CO and C$^{18}$O to the observation program M12AU28.
The angular resolution of the whole set of molecular data is about 14\arcsec.

\section{Results and discussion}

\subsection{Molecular characterization of the BRC} 

\begin{figure}[!t]
    \centering
    \includegraphics[width=0.4\textwidth]{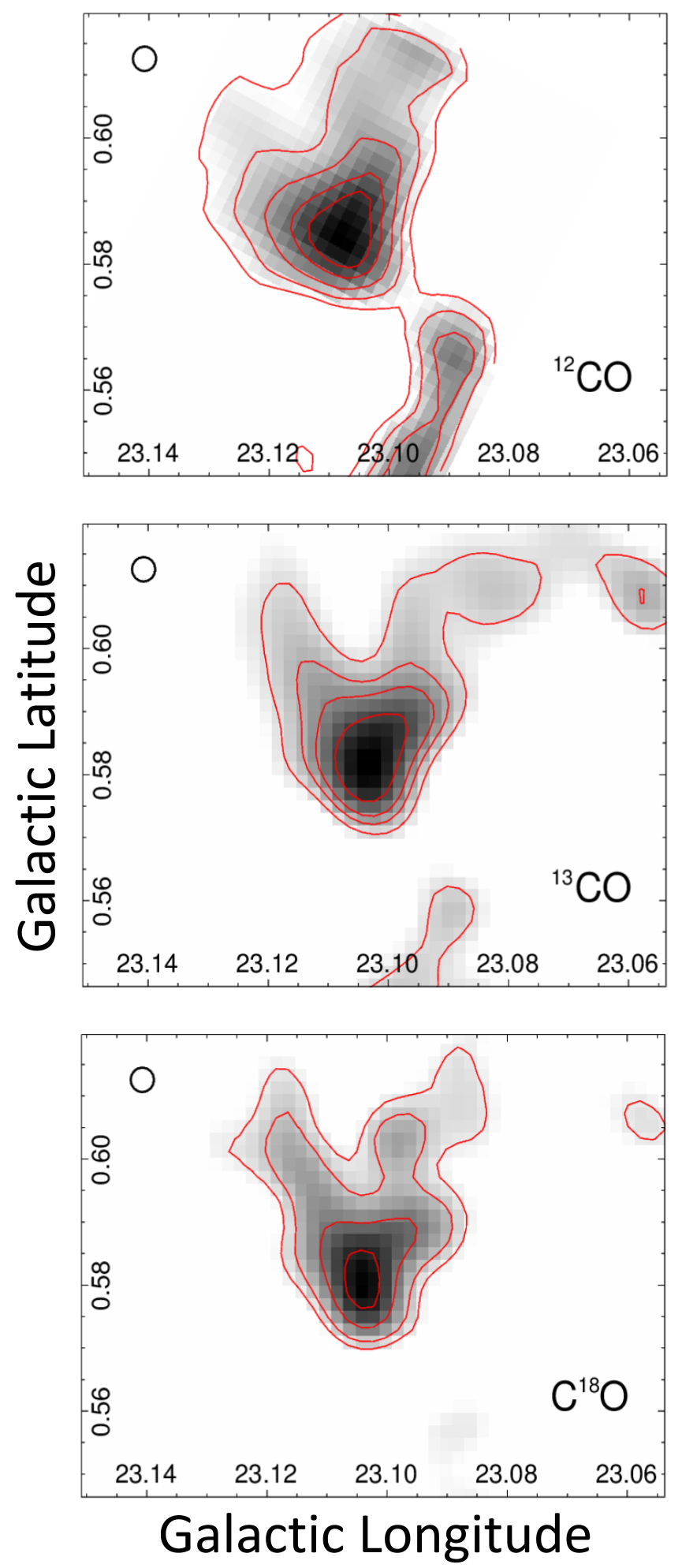}
    \caption{$^{12}$CO, $^{13}$CO and C$^{18}$O J=3--2 integrated maps between 30 and 45 km s$^{-1}$. The beam of the data are included at the top left corner in each panel. The contours levels are: 18, 35, 40, 60 and 66 K km s$^{-1}$ ($^{12}$CO), 8, 12, 15, and 20 K km s$^{-1}$ ($^{13}$CO), and 1.0, 1.5, 2.5 and 3.5 K km s$^{-1}$ (C$^{18}$O) }
    \label{CO}
\end{figure}

\begin{figure}[!t]
    \centering
    \includegraphics[width=0.48\textwidth]{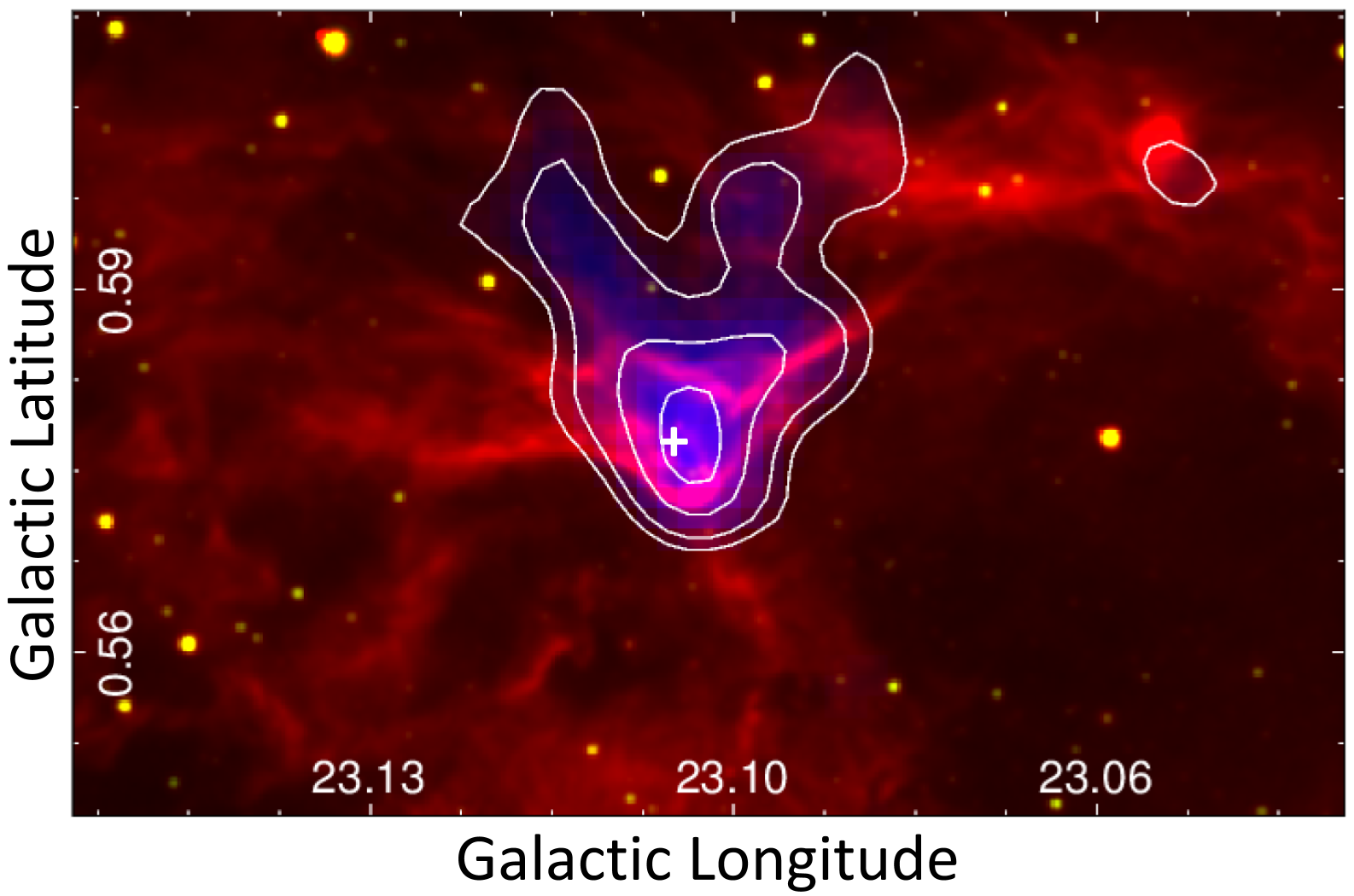}
    \caption{Three colour image displaying the GLIMPSE/{\it Spitzer} emissions at 
    8 and 4.5~$\mu$m (red and green, respectively) with the integrated C$^{18}$O J=3--2 emission (in blue). The contours are the same as presented in Figure\,\ref{CO} bottom panel. The cross indicates the position of the source SSTGLMC G023.1052+00.5816.}
    \label{color}
\end{figure}

We carefully analyzed the data cubes along the whole velocity range of each CO isotope, and we found that the molecular emission related to the BRC is concentrated between 30 and 45  km s$^{-1}$. Figure\,\ref{CO} shows the integrated maps of each isotope along this velocity range. Figure\,\ref{color} displays the C$^{18}$O J=3--2 emission (in blue with white contours) superimposed to the 8 and 4.5 $\mu$m emissions, which 
clearly shows a perfect morphological correspondence between the molecular feature and the BRC.

By assuming local thermodynamic equilibrium (LTE), and following the typical formulae (see for instance \citealt{areal18}) the column density of C$^{18}$O (N(C$^{18}$O)) was estimated towards the BRC. Using the obtained N(C$^{18}$O) and the relation [H$_{2}$]/[$^{18}$O]$=5.8 \times 10^{6}$ \citep{frer82} a
H$_{2}$ column density was obtained. Finally, assuming a distance of about 2.2 kpc and considering the cloud morphology, a molecular mass of about 230 M$_{\odot}$ was derived for the molecular feature associated with the BRC, which has a size of about 1.5 pc. The obtained molecular mass and the size are typical for this kind of structures, and these values are usually used in the simulations of the RDI process in BRCs (e.g. \citealt{fukuda13}).




\subsection{$^{13}$CO/C$^{18}$O abundance ratio} 

\begin{figure}[!t]
    \centering
    \includegraphics[width=0.35\textwidth]{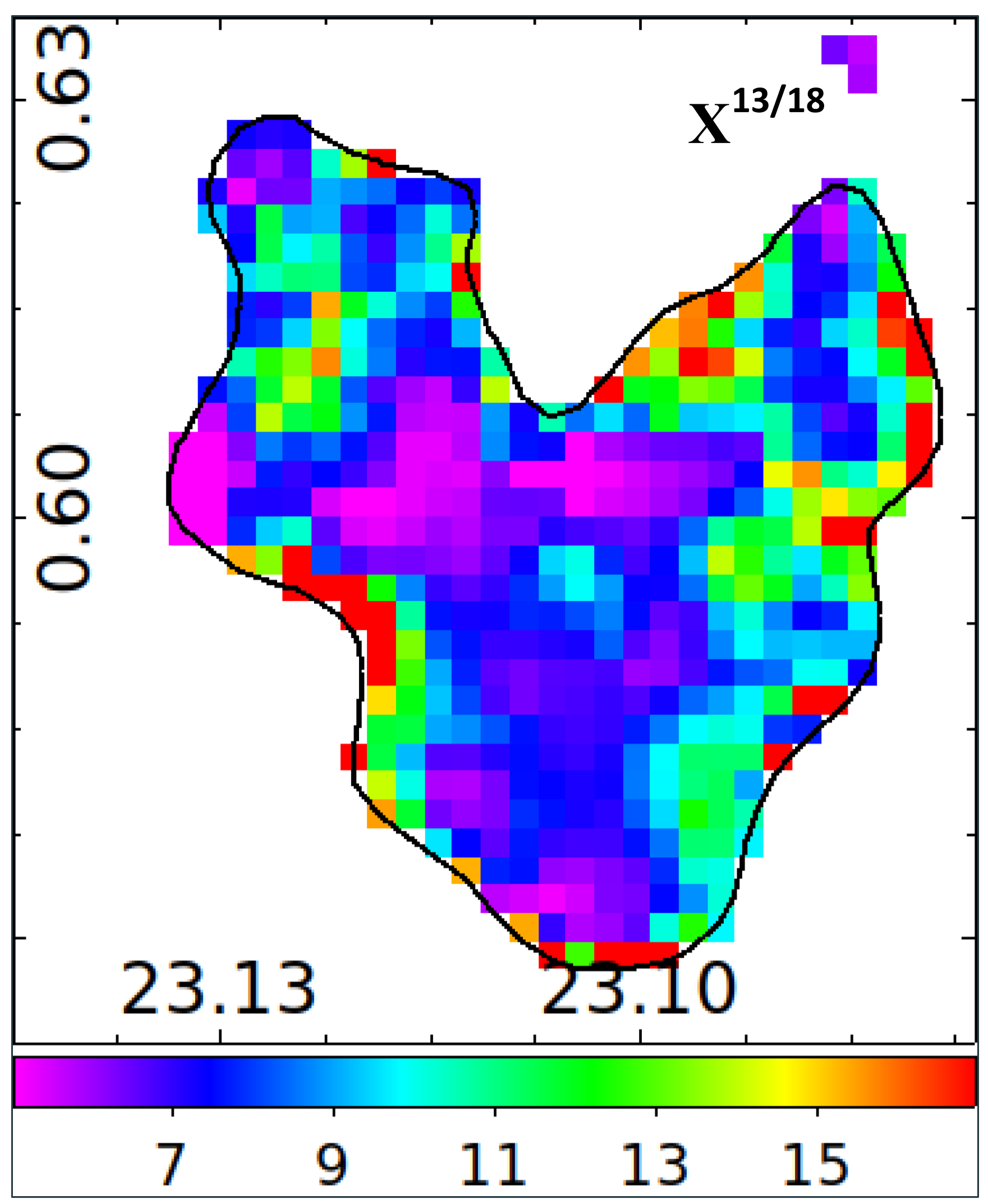}
    \caption{Map of abundance ratio X$^{13/18}$ (N($^{13}$CO)/N(C$^{18}$O)) towards the analyzed BRC at the border of bubble N30.}
    \label{X}
\end{figure}

Given that the $^{13}$CO/C$^{18}$O abundance ratio is an useful parameter to
study the relation between the molecular gas and the far ultraviolet (FUV) radiation (e.g. \citealt{paron18,shima14}),
by assuming LTE as done above (again, see the formulae and the whole procedure in \citealt{areal18}), we calculated the $^{13}$CO and C$^{18}$O column densities
pixel-by-pixel within the molecular feature delimited by the C$^{18}$O emission. The map of the abundance ratio X$^{13/18}$ (N($^{13}$CO)/N(C$^{18}$O)) towards the BRC is presented in Figure\,\ref{X}. It can be seen that this ratio in general increases towards the edges of the molecular structure. This suggests that C$^{18}$O is selectively photodissociated with respect to $^{13}$CO in the regions most intensely irradiated by the FUV radiation from the stars that gave rise to the H{\sc ii} region. It is likely that this radiation, in turn, is shaping the studied molecular structure.

\subsection{Young stellar object candidate in the BRC}

From a catalog search we found a young stellar object (YSO) candidate within the analyzed region. The object is the intrinsically red source observed
by {\it Spitzer} SSTGLMC G023.1052+00.5816, which according to \citet{robi08} could be a YSO. The position of this source is indicated in Figure\,\ref{color} with 
a cross. We can see that its location coincides in projection with the BRC, suggesting that it could have been formed through the RDI star formation process mentioned
in Section\,1. A further analysis about the internal and external pressures in the BRC will determine whether the RDI mechanism is ongoing in the region.

\begin{acknowledgement}
M.B.A. thanks the financial support received to assist to the AAA meeting held in Viedma. M.B.A. is a doctoral fellow of CONICET. This work is part of the grade thesis ({\it licenciatura}) done by A.S. at Universidad de Buenos Aires.

\end{acknowledgement}


\bibliographystyle{baaa}
\small
\bibliography{biblio}
 
\end{document}